\documentclass[12pt]{iopart}
\usepackage{amssymb}
\usepackage{bm}
\usepackage{graphicx}
\usepackage[
pdfauthor={Paul D. Nation},
pdftitle={Steady-state solution methods for open quantum optical systems},
bookmarks=true,colorlinks=true,linkcolor=blue,urlcolor=blue,citecolor=red
]{hyperref}

\begin{document}

\title{Steady-state solution methods for open quantum optical systems}

\author{P. D. Nation}
\ead{\mailto{pnation@korea.ac.kr}}
\address{Department of Physics, Korea University, Seoul 136-713, Korea}

\date{\today}

\begin{abstract}
We discuss the numerical solution methods available when solving for the steady-state density matrix of a time-independent open quantum optical system, where the system operators are expressed in a suitable basis representation as sparse matrices.  In particular, we focus on the difficulties posed by the non-Hermitian structure of the Lindblad super operator, and the numerical techniques designed to mitigate these pitfalls.  In addition, we introduce a doubly iterative inverse-power method that can give reduced memory and runtime requirements in situations where other iterative methods are limited due to poor bandwidth and profile reduction.  The relevant methods are demonstrated on several prototypical quantum optical systems where it is found that iterative methods based on iLU factorization using reverse Cuthill-Mckee ordering tend to outperform other solution techniques in terms of both memory consumption and runtime as the size of the underlying Hilbert space increases.  For eigenvalue solving, Krylov iterations using the stabilized bi-conjugate gradient method outperform generalized minimal residual methods.  In contrast, minimal residual methods work best for solvers based on direct LU decomposition.  This work serves as a guide for solving the steady-state density matrix of an arbitrary quantum optical system, and points to several avenues of future research that will extend the applicability of these classical algorithms in absence of a quantum computer.
\end{abstract}

\submitto{New J. Phys.} 
\maketitle

\section{Introduction}\label{sec:intro}
Understanding the interplay between a quantum system and its environment is of fundamental importance for realistic quantum systems \cite{breuer:2002}.  For although much care is taken experimentally to eliminate the unwanted influence of external interactions, there remains, if ever so slight, a coupling between the system of interest and the external world.  In addition, any measurement performed on the system, classical or quantum, necessarily involves coupling to the measurement device, thereby introducing an additional source of external influence \cite{clerk:2010,wiseman:2010}.  By definition, an open quantum system is coupled to an environment where the complexity of the environmental dynamics renders the combined evolution of system plus reservoir intractable.  However, for a system weakly coupled to its surroundings, there is a clear distinction between the system and its environment, allowing for the dynamics of the latter to be traced over, resulting in a reduced density matrix describing the system alone.

Finding the steady-state solution to the reduced density matrix of an open quantum system is of great importance for both experimental and theoretical investigations of quantum optics and related systems such as trapped ions, superconducting circuits,  and quantum nanomechanical systems (e.g. see Refs.~\cite{grimsmo:2013,majumdar:2013,nation:2013,reiter:2013,chen:2014,hush:2014,killoran:2014,lorch:2014,prado:2014,schlawin:2014,weimer:2014,holland:2015}).  In addition to being a good approximation to the density matrix for many experimental quantum systems, in some devices it is possible to observe quantum features in the steady-state density matrix, such as negative Wigner functions \cite{lorch:2014, nation:2015, holland:2015}, under experimentally feasible operating conditions that can be repeatedly measured without detrimentally affecting the state of the system.  Often the steady-state density matrix cannot be expressed analytically, and as such, the efficient numerical simulation of this operator is a key component in the exploration of a multitude of quantum optical systems.

In absence of a quantum computer, the numerical calculation of the steady-state density matrix must be solved using classical computing resources, where the fundamental limit on the size of quantum system that can be explored is constrained by the exponentially increasing dimensionality of the underlying composite Hilbert space.  However, when representing the quantum mechanical operators in a chosen basis using sparse matrices, the dimensionality of the underlying Hilbert space is currently not the limiting factor in determining the steady-state solution for a quantum optical system.  Rather, it is the difficulties that arise due to the lack of Hermicity and poor conditioning of the Liouvillian super operator that lead to large runtimes and memory consumption when using sparse linear solution methods.  As such, a thorough understanding of the available algorithms, and the development of techniques designed to overcome these limitations, is crucial to enabling the classical simulation of ever larger quantum systems.

In this work, we detail the numerical solution methods available when solving for the steady-state density matrix for an arbitrary time-independent quantum optical system where the operators are represented as  sparse matrices.  We consider the standard eigenvalue and linear system solution methods, along with recently developed iterative techniques and bandwidth reducing permutation methods designed to overcome the limitations of these standard approaches \cite{nation:2015}.  In addition, we put forth a doubly-iterative inverse power method, that yields better performance as compared to previously considered iterative methods when bandwidth reduction permutations are less effective.  Here we show that, with the addition of this new doubly-iterative algorithm, iterative solution methods, in general, outperform their direct factorization counterparts in terms of both algorithm runtime and memory consumption for large quantum systems provided that the Liouvillian operator is suitably permuted such that the approximate inverse is well-conditioned.  Both memory and solution time benchmarks can be reduced by an order of magnitude, or more, when using the appropriate iterative technique for a given Liouvillian.  These methods allow for the simulation of ever larger quantum systems, and help to mitigate the need for a quantum computer.

This paper is organized as follows.  In Sec.~\ref{sec:steady} we introduce the eigenvalue equation whose solution is the steady-state density matrix for which we must solve. Section~\ref{sec:eigen} explores the use of eigenvector solution methods, their dependence on LU factorization, and the requirement of well-conditioned eigenvalues.  In Sec.~\ref{sec:direct} we make use of the unit trace property of the density matrix to recast the eigenvalue problem as a linear system of equations, and discuss matrix permutation techniques designed to reduce fill-in in the LU decomposition that affects both memory and runtime.  Section~\ref{sec:iterative} discusses preconditioned iterative solvers and highlights the difficulties in implementing these techniques as general solution methods.  Numerical simulations of several canonical quantum optical systems implementing these solution methods are presented in Sec.~\ref{sec:numerics}.  Finally, Sec.~\ref{sec:conclusion} concludes with a brief discussion of the results and directions for possible future research.

\section{Steady-state density matrix}\label{sec:steady}
For an open quantum system with decay rates larger than the corresponding excitation rates, the system approaches the steady-state density matrix $\hat{\rho}_{\rm ss}$ as $t\rightarrow \infty$ satisfying the eigenvalue equation
\begin{equation}\label{eq:ss}
\frac{d \hat{\rho}_{\rm ss}}{dt}=\mathcal{L}\left[\hat{\rho}_{\rm ss}\right]=0,
\end{equation}
where $\mathcal{L}$ is the Liouvillian super operator.  In many quantum optical systems, the Liouvillian takes the most general Lindblad form
\begin{equation}\label{eq:lindblad}
\mathcal{L}[\hat{\rho}]=-\frac{i}{\hbar}[\hat{H},\hat{\rho}]+\sum_{k}\mathcal{D}[\hat{C}_{k},\hat{\rho}]
\end{equation}
with the dissipative terms given by
\begin{equation}
\mathcal{D}[\hat{C}_{k},\hat{\rho}]=\frac{1}{2}[2\hat{C}_{k}\hat{\rho}\hat{C}^{\dag}_{k}-\hat{\rho}\hat{C}^{\dag}_{k}\hat{C}_{k}-\hat{C}^{\dag}_{k}\hat{C}_{k}\hat{\rho}],
\end{equation}
where the $\hat{C}_{k}=\sqrt{\gamma_{k}}\hat{A}_{k}$ are collapse operators determined by the rates $\gamma_{k}$ and operators $\hat{A}_{k}$ for each dissipation channel that couples the system to its environment or measurement apparatus.  This form of the Liouvillian is not required for the analysis presented here, and can be substituted with any time-independent Liouvillian.

If the system Hamiltonian and collapse operators are time-independent, or can be transformed into such a form by moving to an interaction representation, then Eq.~(\ref{eq:ss}) can be recast as a sparse matrix eigenvalue equation
\begin{equation}\label{eq:eigval}
\bm{\mathcal{L}}\vec{\rho}_{\rm ss}=0\vec{\rho}_{\rm ss},
\end{equation}
where $\vec{\rho}_{\rm ss}$ is the dense vector formed by vectorization (column stacking) of $\hat{\rho}_{\rm ss}$, and $\bm{\mathcal{L}}$ is the sparse matrix representation of the Liouvillian in a given basis representation.  In what follows, we will always assume a Fock state basis for oscillator modes, and a $z$-basis representation for spin operators.  Given the non-Hermitian form of the Liouvillian, the solutions to Eq.~(\ref{eq:eigval}) are non-trivial, and the methods by which this equations can be solved is the focus of the remainder of this work.

\section{Eigenvalue Methods}\label{sec:eigen}

Let us assume that the eigenvalues of the Lindblad operator $\bm{\mathcal{L}}\in \mathbb{C}^{n}\times \mathbb{C}^{n}$, formed from a composite Hilbert space with dimensionality $\rm{dim}~\mathcal{H}=\sqrt{n}$, can be diagonalized with eigenvalues ordered as $|\lambda_{1}|\ge|\lambda_{2}|\ge\dots\ge|\lambda_{n}|$.  Using the spectral decomposition $\bm{\mathcal{L}}=\bm{\mathcal{V}}\bm{\Lambda} \bm{\mathcal{V}}^{-1}$ where $\bm{\Lambda}=\rm{diag}(\lambda_{1},\dots,\lambda_{n})$ and $\bm{\mathcal{V}}$ is the matrix with the eigenvectors of  $\bm{\mathcal{L}}$ as the column entries, repeated application of the Liouvillian obeys the relationship

\begin{equation}\label{eq:powerid}
\bm{\mathcal{L}}^{k}=\bm{\mathcal{V}}\bm{\Lambda}^{k}\bm{\mathcal{V}}^{-1}. 
\end{equation}

For the vector $\vec{x}_{0}=\bm{\mathcal{V}}\vec{\tilde{x}}\in \mathbb{C}^{n}$, generated from a random vector $\vec{\tilde x}$, the application of the Lindblad operator $k$-times can be expressed as
\begin{equation}\label{eq:kpower}
\bm{\mathcal{L}}^{k}\vec{x}_{0}=\lambda^{k}_{1}\left[\sum_{i=1}^{n}v_{i}\left(\frac{\lambda_{i}}{\lambda_{1}}\right)^{k}\vec{\tilde{x}}\right],
\end{equation} 
with $(\lambda_{i}/\lambda_{1})^{k}\rightarrow 0$ for $i > 1$ and a sufficiently large $k$ provided that $\vec{\tilde{x}}_{1}\neq 0$, i.e. the initial vector has a nonzero component in the direction of the dominate eigenvalue, and $\lambda_{1}$ is well separated from $\lambda_{2}$ .  The former criterion is automatically met by starting with a random vector.  Provided that these conditions are satisfied then $\bm{\mathcal{L}}^{k}\vec{x}$ is approximately parallel to the dominate eigenvector $v_{1}$ with error decreasing linearly at an asymptotic rate given by $|\lambda_{2}|/|\lambda_{1}|$.  The power series solution for the dominate eigenvector can therefore be written as
\begin{equation}\label{eq:power}
\vec{x}_{k+1}=\frac{A\vec{x}_{k}}{||A\vec{x}_{k}||}=\frac{A^{k}\vec{x}_{0}}{||A^{k}\vec{x}_{0}||},
\end{equation}
where $||\cdot||$ is a chosen vector norm, e.g. the $L2$-norm. Since any eigenvalue problem is equivalent to finding the roots of the characteristic polynomial, from the Abel-Ruffini theorem, it is impossible to directly compute the eigenvalues for any matrix with dimension $\ge5$ \cite{trefethen:1997}.  Power iteration itself is not suitable for finding the steady-state density matrix as the zero eigenvalue is not the dominate eigenvalue of $\bm{\mathcal{L}}$.  Furthermore, if $|\lambda_{2}|/|\lambda_{1}|\sim 1$, then the convergence of this method is prohibitively slow.  The former of these shortcomings can be addressed by modifying the power iteration method, whereas the latter is not encountered in practice. 

To transform the power method into an algorithm suitable for our purposes, recall that this method relies on the identity given in Eq.~(\ref{eq:powerid}).  Given a function $f(z)$, defined locally via a convergent power series,  with the eigenvalues of $\bm{\mathcal{L}}$ in the radius of convergence, then one can define $f\left(\bm{\mathcal{L}}\right)$ via the same series, and $f\left(\bm{\mathcal{L}}\right)=\bm{\mathcal{V}}f\left(\bm{\Lambda}\right)\bm{\mathcal{V}}^{-1}$ with $f\left(\bm{\Lambda}\right)=\rm{diag}[f(\lambda_{1}),\dots,f(\lambda_{n})]$.  For the specific choice of $f(z)=(z-\sigma)^{-1}$ one has
\begin{equation}\label{eq:inverse}
\left(\overrightarrow{\bm{\mathcal{L}}}\right)^{-1}=\left(\bm{\mathcal{L}}-\sigma \bm{\mathcal{I}}\right)^{-1}=\bm{\mathcal{V}}\left(\bm{\Lambda}-\sigma\bm{\mathcal{I}}\right)^{-1}\bm{\mathcal{V}}^{-1},
\end{equation}
where $\bm{\mathcal{I}}$ is the identity matrix and the shifted-Liouvillian is denoted by $\overrightarrow{\bm{\mathcal{L}}}$.

Performing power iteration on Eq.~(\ref{eq:inverse}) converges to the dominate eigenvalue $\lambda_{j}$  where $(\lambda_{j}-\sigma)^{-1}$ is maximal; the solution vector corresponds to the eigenvalue in the complex plane closest to $\sigma$.  This method is known as the Shifted-Inverse Power technique \cite{trefethen:1997} where the iterations are  given by
\begin{equation}\label{eq:inverseiter}
\vec{x}_{k+1}=\frac{\left(\overrightarrow{\bm{\mathcal{L}}}\right)^{-1}\vec{x}_{k}}{\left|\left|\left(\overrightarrow{\bm{\mathcal{L}}}\right)^{-1}\vec{x}_{k}\right|\right|}.
\end{equation}
Although Eq.~(\ref{eq:inverseiter}) is formally correct, in practice, it is both computationally and memory intensive to calculate the inverse of a sparse matrix, as the inverse is in general dense.  However, the product $\left(\overrightarrow{\bm{\mathcal{L}}}\right)^{-1}\vec{x}_{k}$ is equivalent to solving the linear system $\overrightarrow{\bm{\mathcal{L}}}\vec{v}=\vec{x}_{k}$, and thus the inverse-power method requires computing the direct sparse LU factorization of the matrix $\overrightarrow{\bm{\mathcal{L}}}$.  This factorization need only be done once, with the solution for different $\vec{x}_{k}$ requiring only computationally efficient forward and backward substitution.  The convergence of shifted-inverse iteration is always linear in the number of iterations.  However, the rate of convergence is determined by the accuracy in choosing $\sigma$ and the proximity of the dominant eigenvalue to its nearest neighbor.  As we know the exact value of the desired eigenvalue, $\sigma=0$, and the dominant eigenvalue is well separated, the number of iterations required to reach a given error tolerance is quite small, usually requiring only a single iteration to reach convergence for a random input vector.  This holds true even when converging to machine precision.  As such, the computational and memory intensive portion of inverse-power iteration lies in the LU factorization of the shifted Liouvillian operator.

While Eq.~(\ref{eq:eigval}) requires finding the eigenvector associated with the zero eigenvalue, in practice, the choice of $\sigma$ is often nonzero and small.  When working with double-precision numbers, a value of $\sigma=10^{-15}$ is typically used.  The effect of this slight shift is to guarantee that the matrix has a nonzero main diagonal, while simultaneously perturbing the matrix away from being strictly singular.  Such a transformation is crucial to the successful reduction in memory consumption for sparse LU factorization provided by the sparse reordering methods discussed in Sec.~\ref{sec:direct} where a nonzero diagonal is assumed.  

Given that $\overrightarrow{\bm{\mathcal{L}}}$ is nearly singular, this matrix is extremely ill-conditioned, and one would not expect an accurate result for the the steady-state density matrix.  However, it can readily be shown that the effectiveness of inverse iteration does not depend on the condition number of $\overrightarrow{\bm{\mathcal{L}}}$ \cite{trefethen:1997}.  Rather it is the condition number of the zero eigenvalue and its eigenvector that determines the stability of the solution.  For a non-degenerate eigenvalue, the associated condition number gives a measure of how close to degenerate the eigenvalue becomes for a given permutation of the system matrix. Unlike Hermitian matrices, where the eigenvalue spectrum is stable against such perturbations \cite{trefethen:1997}, the eigenvalues of non-symmetric matrices can be arbitrarily ill-conditioned \cite{saad:1989}, and the associated loss of accuracy can give rise to considerable errors in the computed eigenvectors.  As the convergence rate of inverse power iteration also depends on the eigenvalue of interest being well separated, poorly conditioned eigenpairs lead to poor performance when using this algorithm.  Prior to computing the eigenvalues, applying a similarity transform $\bm{\mathcal{D}}\bm{\mathcal{L}}\bm{\mathcal{D}}^{-1}$, where $\bm{\mathcal{D}}$ is a diagonal matrix, that reduces the norm of $\bm{\mathcal{L}}$, or the condition number of a subset of eigenvalues can be advantageous \cite{chen:2000}.  The possibility of ill-conditioned eigenvalues must be kept in mind when using eigenvalue solving methods, but is rarely encountered in practice \footnote{We are unaware of any example where the conditioning of the zero eigenvalue leads to a break down in the inverse-power method.  However, ill-conditioning of eigenvalues can give spurious results when computing the general eigenvalues of a Liouvillian operator \cite{grimsmo:2013b}.}.  Indeed, the inverse power method typically converges in a single iteration.  Supplementing the numerical solution with analytical results, and/or varying parameter values over a given range and looking for discontinuous jumps in quantities such as density matrix metrics, for example the fidelity or trace distance, and expectation values can help in part help to diagnose the presence of an ill-conditioned zero eigenvalue.

Finally, Eq.~(\ref{eq:eigval}) can also be solved using algorithms suitable for finding a subset of eigenvalues of a sparse matrix such as, for example, ARPACK \cite{arpack:1997}.  This method works by building the Krylov subspace formed from repeated applications of the shifted Liouvillian on the initial random vector $\vec{x}_{0}$
\begin{equation}\label{eq:krylov}
\mathcal{K}_{n} = \textrm{span}\left\{\vec{x}_{0},A\vec{x}_{0},A^{2}\vec{x}_{0},\dots,A^{n-1}\vec{x}_{0}\right\},
\end{equation}
with $A=\left(\overrightarrow{\bm{\mathcal{L}}}\right)^{-1}$, followed by an orthogonalization procedure based on Arnoldi iteration.   For a non-symmetric matrix, building the Krylov subspace requires first performing the sparse LU factorization of $\overrightarrow{\bm{\mathcal{L}}}$ and, as in the inverse-power method, this factorization represents the bulk of the runtime and memory usage in this method.

\section{Direct factorization}\label{sec:direct}
 As an alternative to solving for the zero eigenvalue, it is possible to find a direct (i.e non-iterative) solution to $\vec{\rho}_{\rm ss}$ via sparse LU decomposition \cite{davis:2006} by making use of the unit trace property of the density matrix to recast the eigenvalue equation as a linear system of equations
\begin{equation}\label{eq:linear}
\tilde{\bm{\mathcal{L}}}\vec{\rho}_{\rm ss}=\left[\bm{\mathcal{L}}+w\bm{\mathcal{T}}\right]\vec{\rho}_{\rm ss}=\left(\begin{array}{c}w \\ 0 \\ \vdots\end{array}\right); \ \ \bm{\mathcal{T}}\vec{\rho}_{\rm ss}=\left(\begin{array}{c}\mathrm{Tr}\hat{\rho}_{\rm ss} \\ 0 \\ \vdots\end{array}\right),
\end{equation}
where $w$ is a controllable weighting factor and $\bm{\mathcal{T}}$ is a matrix that enforces $\mathrm{Tr}\left(\hat{\rho}_{\rm ss}\right)=1$, with ones along the upper row in the columns corresponding to the locations of the diagonal elements in $\vec{\rho}_{\rm ss}$.  Although the choice of row for the nonzero values in $\bm{\mathcal{T}}$ is arbitrary, importantly, this matrix always has a nonzero element in the final column.  In contrast to iterative methods, the eigenvalue condition number has no effect on direct factorization, making this method attractive for finding steady-state solutions to open quantum systems. Note that the restriction $\mathrm{Tr}\left(\hat{\rho}_{ss}\right)=1$ is already included in the Liouvillian operator $\mathcal{L}$, and its matrix representation $\bm{\mathcal{L}}$.  Here, this constraint is used to simply add a constant vector to both sides of Eq.~(\ref{eq:eigval}).

When performing the LU factorization of a sparse matrix, nonzero elements arise in the $L$ and $U$ factors that are not present in the original matrix; the sparse structure of $LU$ is not the same as $\tilde{\bm{\mathcal{L}}}$ \cite{davis:2006}.  This fill-in, must be minimized in order to reduce both the memory requirements for storing the LU factors and the runtime of factorization, both of which scale with the number of nonzero matrix elements $\rm NNZ$ \cite{davis:2006}. A convenient measure of the fill-in is given by the ratio of nonzero elements in the $L$ and $U$ factors to those in the original Liouvillian $(L_{\rm NNZ}+U_{\rm NNZ})/\tilde{\bm{\mathcal{L}}}_{\rm NNZ}$ called the fill factor. The fill-in is sensitive to the order in which the rows and columns of a sparse matrix are operated on, and in particular, to the matrix bandwidth size and profile \cite{george:1981}.  For a non-symmetric sparse matrix $\bm{A}=\{a_{ij}\}$, one can define the upper and lower bandwidths, `$\rm ub$' and `$\rm lb$' respectively, to be
\begin{equation}
\rm{ub}=\max_{a_{ij}\neq0}(j-i); \ \ \rm{lb}=\max_{a_{ij}\neq0}(i-j).
\end{equation}
The total bandwidth $B$ is then the sum $B=\rm{ub}+\rm{lb}+1$, where the one takes into account the main diagonal \cite{saad:2003}.  Likewise, we define the upper profile $\rm up$ and lower profile $\rm lp$ as
\begin{equation}
\rm{up}=\sum_{i} \max_{a_{ij}\neq0}(j-i); \ \ lp=\sum_{j} \max_{a_{ij}\neq0}(i-j),
\end{equation}
such that the total profile is $P=\rm{up}+\rm{lp}$.  From the definition of the bandwidth, we see that imposing the trace condition with the matrix $\bm{\mathcal{T}}$ gives an upper bandwidth that is equal to the square of the dimensionality of the Hilbert space.  Therefore, the total bandwidth is bounded below by this value, suggesting that the fill-in for the LU factors rises rapidly with system size.  Arising from the use of the trace matrix $\bm{\mathcal{T}}$, and not the form of the Liouvillian, this bandwidth scaling holds for any system.  It is therefore advantageous to find a matrix permutation that moves the elements of $\bm{\mathcal{T}}$, as well as all other elements, toward the diagonal, thereby reducing the overall bandwidth and profle of the modified Liouvillian.

Finding the minimal bandwidth of a matrix is an NP-complete problem \cite{yannakakis:1981}, and therefore several heuristics have been developed that attempt to minimize the bandwidth while remaining computationally efficient.  One common technique is to permute the rows and columns of a symmetric matrix based on the Cuthill-McKee (CM) ordering \cite{cuthill:1969}.  Taking the structure of a symmetric sparse matrix as the adjacency matrix of a graph, CM ordering does a breadth-first search (BFS) of the graph starting with a node (row) of lowest degree, where the degree of the $i^{th}$ node is defined to be the number of nonzero elements in the $i^{th}$ matrix row, and visiting neighboring nodes in each level-set in order from lowest to highest degree.  This is repeated for each connected component of the graph.  CM ordering also reduces the profile of the matrix, and it was noticed that reversing the CM order, the RCM ordering, gives a superior profile reduction while leaving the bandwidth unchanged \cite{george:1981}.  Since RCM operates on the structure of a matrix, only the locations of nonzero matrix elements, and not their numerical values, are used in this reordering.  Modified RCM methods, such as the Gibbs-Poole-Stockmeyer (GPS) algorithm \cite{gibbs:1976}, provide a similar bandwidth and profile reduction. In the Fock basis, the effect of RCM reordering is to permute the basis vectors such that the Fock states are no longer in ascending order.  

As the Liouvillian operator is itself non-symmetric, we  calculate the RCM ordering of the symmetrized form $\tilde{\bm{\mathcal{L}}}+\tilde{\bm{\mathcal{L}}}^{T}$, and apply the resulting row and column ordering to $\tilde{\bm{\mathcal{L}}}$ to obtain $\tilde{\bm{\mathcal{L}}}_{\rm RCM}$. One can also used the symmetrized product $\tilde{\bm{\mathcal{L}}}\tilde{\bm{\mathcal{L}}}^{T}$ or $\tilde{\bm{\mathcal{L}}}^{T}\tilde{\bm{\mathcal{L}}}$, however this would lead to dense matrices and large memory consumption unless done symbollically.  While RCM reordering of a symmetric matrix is guaranteed to reduce the bandwidth and profile, or at least do no worse \cite{george:1981}, the need to operate on $\tilde{\bm{\mathcal{L}}}+\tilde{\bm{\mathcal{L}}}^{T}$ rather than $\tilde{\bm{\mathcal{L}}}$ itself invalidates this assurance. RCM reordering of $\tilde{\bm{\mathcal{L}}}+\tilde{\bm{\mathcal{L}}}^{T}$ overestimates the graph structure of the original modified Liouvillian, and there is little a priori information from which to judge whether RCM will significantly reduce the bandwidth and profile when the structure of the Liouvillian operator is varied.  This is especially true when the structure of $\tilde{\bm{\mathcal{L}}}$ is far from symmetric, and the two matrix structures differ by a significant amount.  Structural changes such as setting to zero numerical system parameters in the Hamiltonian or collapse operators can also affect the performance of RCM ordering. 

Along with symmetric reorderings that reduce both the bandwidth and profile of the Liouvillian, it is possible to reduce the fill-in via non-symmetric permutations that sort only the columns, or rows, of the matrix.  In the present case of non-symmetric matrices, the best performing general purpose reordering strategy is the Column Approximate Minimum Degree (COLAMD) ordering \cite{davis:2004} that is the default reordering method in the SuperLU linear solver used here \cite{demmel:1999}, as well as in many commercial solvers such as those found in Matlab \cite{matlab}.  This method finds the symmetric permutation of $\tilde{\bm{\mathcal{L}}}^{T}\tilde{\bm{\mathcal{L}}}$ that, when applied as a column permutation to $\tilde{\bm{\mathcal{L}}}$, minimizes the worst case fill factor for an arbitrary permutation of the matrix rows.  In this case, the product $\tilde{\bm{\mathcal{L}}}^{T}\tilde{\bm{\mathcal{L}}}$ is computed symbolically to avoid the explicit construction of the resulting dense matrix.

Minimizing the fill-in using COLAMD reordering requires a matrix with a nonzero main diagonal.  In the case of the inverse-power method, this condition was automatically satisfied when applying the eigenvalue shift. In contrast, the diagonal of the modified Liouvillian can, in general, contain zero elements that must be permuted away.  As the modified Liouvillian is necessarily non-singular, it is always possible to find a non-symmetric permutation that makes the diagonal zero-free \cite{davis:2006}.  A simple method, using only the matrix structure, is Maximum Bipartite Matching (MBM) \cite{duff:2011}.  Here we implementation a variation of the MBM algorithm using a BFS method processing the columns of the matrix in order starting with the column possessing the largest absolute value element.  From this initial column, the algorithm proceeds via a BFS where the rows are processed in descending order based on the absolute value of each row element. The corresponding columns are marked as being visited, and the routine continues until all columns have been processed. The goal of this weighted variant of the MBM algorithm is to permute the matrix to a zero-free diagonal while simultaneously attempting to increase the sum of the diagonal elements.  This can help, in part, to overcome difficulties in the construction of approximate LU factorizations discussed in Sec.~\ref{sec:iterative} \cite{chow:1997} . Although the diagonal sum is not guaranteed to be maximized, this method is computationally less intensive than maximum product traversal \cite{duff:1999} and related methods.  Note that RCM permutation does not require a non-zero diagonal, and in many cases, applying a non-symmetric permutation such as MBM before RCM can lead to both a larger bandwidth and profile as the matrix structure becomes less symmetric.

\section{Iterative Solvers}\label{sec:iterative}
The main drawback in using direct LU decomposition is the poor scaling in terms of both runtime and memory requirements, even when symmetric and/or non-symmetric reordering methods are employed, as the dimension of the matrix increases.  Therefore, for sufficiently large sparse matrices, iterative methods are the only available solution method \cite{benzi:2002}, with the most common choice being iterative Krylov solvers \cite{saad:2003}.  While iterative methods require less memory and fewer numerical operations than direct methods, these methods usually require preconditioning to achieve a sufficient tolerance level and a reasonable convergence rate \cite{benzi:2002}.  The goal of preconditioning is to convert the original system of equations, Eq.~(\ref{eq:linear}), into the modified linear system 
\begin{equation}\label{eq:modified}
\bm{\mathcal{M}}^{-1}\tilde{\bm{\mathcal{L}}}\vec{\rho}_{ss}=\bm{\mathcal{M}}^{-1}\left(\begin{array}{c}w \\ 0 \\ \vdots\end{array}\right),
\end{equation}
where $\bm{\mathcal{M}}$ is the preconditoner.  Convergence is improved provided that the condition number of $\bm{\mathcal{M}}^{-1}\tilde{\bm{\mathcal{L}}}$ is significantly lower than that of $\tilde{\bm{\mathcal{L}}}$ itself.  As the condition number of a sparse matrix grows with the dimensionality \cite{davis:2006}, preconditioning is required when solving large sparse systems.  The best preconditioner is obviously $\bm{\mathcal{M}}=\tilde{\bm{\mathcal{L}}}$, however this is equivalent to solving the original system of equations.  Instead, it is possible to efficiently solve for an approximation of the  inverse $\bm{\mathcal{M}}\approx\tilde{\bm{\mathcal{L}}}$.  The application of a suitable preconditioner should make the modified linear system Eq.~(\ref{eq:modified}) easy to solve, while the preconditioner itself should be simple to build and apply as one or more matrix-vector products are required for each iteration.  Moreover, the condition number of $\bm{\mathcal{M}}$ should not be so large as to affect convergence.

The iLU class of preconditoners are constructed from an incomplete (approximate) LU factorization to the modified Liouvillian $\tilde{\bm{\mathcal{L}}}$ by discarding fill-in elements based on a designated dropping strategy.  The method used here is an incomplete LU factorization with dual-threshold and pivoting (iLUTP) \cite{saad:2003}, where a drop tolerance $d$ and allowed fill-in $p$ are specified such that all fill-in smaller than $d$ times the infinity-norm of a row are dropped, and at most only $p\cdot \tilde{\bm{\mathcal{L}}}_{\rm{NNZ}}$ fill-in elements are allowed.  Note that these parameters are not independent.  In the limit where $d=0$, the iLU preconditioner returns the complete LU factorization, and the fill-in for the direct factorization can be viewed as an upper-bound on the size of the preconditioner.  The condition number of $\bm{\mathcal{M}}$ can vary as a function of the drop tolerance, and therefore decreasing $d$ does not necessarily improve the convergence rate \cite{chow:1997}.  A convenient estimate for the condition number of the approximate inverse is given by  $||\bm{\mathcal{M}}\vec{e}||_{\infty}$, where $\vec{e}=(1,1,\dots)^{T}$ \cite{chow:1997} and $||\cdot||_{\infty}$ is the infinity-norm.  Although this measure gives only a lower bound on $||\bm{\mathcal{M}}||_{\infty}$, it has proved to be an useful benchmark for the stability of $\bm{\mathcal{M}}$ \cite{nation:2015}. Finding the best combination of parameters for a given matrix is a trial and error process, thus preventing the use of these methods as a general purpose solver.

While the generation of robust iLU preconditioners in the case of an Hermitian matrix is now well established, the existence and stability of preconditioners for non-symmetric matrices is understood to a lesser extent \cite{benzi:2002}.  In the non-symmetric case, iLU preconditioners typically fail due to a lack of diagonal dominance, zeros along the diagonal,  and inaccuracies in the approximate inverse due to the dropping of small nonzero elements \cite{chow:1997}.  Moreover, even if a preconditioner is found, its condition number can be larger than that of the original matrix and hence convergence is lost.  However, studies have shown that these failures may be overcome by utilizing symmetric and/or non-symmetric reorderings of the matrix to maximize the sum of the diagonal elements, and reduce the overall bandwidth and profile \cite{chow:1997, benzi:1999, benzi:2002}.  The majority of these reordering strategies are developed for matrices with symmetric structure (graphs) and therefore their application to non-symmetric problems with differing matrix structures must be evaluated on a case by case basis (see Ref.~\cite{benzi:2002} and references therein); there is no universally applicable reordering scheme for non-symmetric matrices.  However, previous studies on symmetric \cite{bridson:2000} and non-symmetric sparse matrices \cite{benzi:1999} have found that RCM reordering improves iLU conditioning at the expense of a marginal increase in fill-in when compared with other reordering strategies.  An intuitive explanation for the case of symmetric matrices is given in Ref.~\cite{bridson:2000}.  In contrast, some of the best methods for fill-in reduction such as COLAMD do poorly when applied to iLU factorization \cite{benzi:2002}.

Having successfully found a preconditioner for a given drop tolerance $d$ and fill-in $p$, a solution to the steady-state density matrix can, in principle, be found using iterative Krylov solution methods \cite{saad:2003} designed for non-symmetric matrices.  Here, we perform the preconditioned iterations using both restarted generalized minimum residual (GMRES) \cite{saad:1986} and stabilized biconjugate gradient (BiCGSTAB) solvers \cite{vorst:1992}.  The GMRES algorithm constructs the Krylov subspace Eq.~(\ref{eq:krylov}) for the preconditioned system and finds the vector in this subspace with minimal residual, as measured by the $L2$-norm, via Arnoldi iteration.  Unlike the symmetric case, where there is a simple three-step recurrence relation for finding orthogonal Krylov vectors \cite{barrett:1993}, the orthogonalization procedure must be done explicitly, which is both computationally and memory intensive, growing as $\mathcal{O}(m^{2})$ where $m$ is the iteration count.  As such, a restarted variant of the original GMRES method is typically used, where the approximate solution vector after $m$-iterations is calculated in the subspace $\mathcal{K}_{m}$, where $m$ is much less than the dimensionality of the system in question, and then is used as the new initial condition for the next $m$-iterations.  An optimal value for $m$, minimizing excessive work, while still leading to convergence is problem specific and must be found through trial-and-error.  The BiCGSTAB algorithm can be viewed as a biconjugate-gradient step followed by a GMRES step with $m=1$ giving a local reduction in the residual vector, leading to smoother (stabilized) convergence behavior \cite{barrett:1993}.  This method is less computationally intensive than GMRES, but comes at the cost of, in general, poorer convergence properties.  

Regardless of which method is used, an acceptable tolerance on the $L2$-norm of the calculated residual vector must be supplied as a stopping criterion.  In addition, one can specify a maximum number of iterations to perform, before convergence is assumed to be not possible.  Setting strict tolerance values, e.g. $10^{-15}$ requires building accurate, well conditioned preconditioners and possibly many iterations, thus increasing the computation time and memory storage requirements.  In contrast,  allowing more moderate values for the solution tolerance, can give a significant performance benefit in some cases.  For a given solution tolerance, the optimal values for the preconditioner fill factor $p$ and drop tolerance $d$ can vary in a non-trivial manner and must be found empirically.

Here we are interested in using these preconditioned iterative methods in the solution for the modified Liouvillian Eq.~(\ref{eq:linear}), as well as in the LU factorization of $\overrightarrow{\bm{\mathcal{L}}}$ used in the inverse power method.  Given that the inverse-power method itself is an iterative method, we designate this method as the doubly-iterative inverse-power method to distinguish it from the usual inverse-power technique using a direct LU factorization of $\overrightarrow{\bm{\mathcal{L}}}$.  Although using an iterative solver in the inverse-power method is typically not ideal due to the need to solve for a new right-hand vector at each iteration, our method relies on the knowledge that only a single inverse-power iteration is typically needed when solving the density matrix for a given $\overrightarrow{\bm{\mathcal{L}}}$.  Therefore, it is possible to obtain a performance improvement provided that the system is large enough, and one forms a high-quality preconditioner.  In order for this doubly-iterative method to converge, the tolerance on the iterative iLU solution vector must be stricter than the tolerance used to stop the inverse-power iterations. In what follows, we will always set the tolerance on the iterative iLU solver to be an order of magnitude lower than that of the inverse-power iterations.

\section{Numerical Simulations}\label{sec:numerics}
Given the non-Hermitian structure of the Liouvillian operator, the optimal solution method for finding the steady-state density matrix of the system will, in general, be problem specific \cite{benzi:2002}.  Therefore, in order to gauge the general performance of each solution method and matrix reordering methods, we apply both eigenvalue and direct solvers to the steady-state solution for a set of common quantum optical systems.  Note that there is an ambiguity when building the sparse matrix representation of a given system Liouvillian since the tensor product, and therefore matrix structure, depends on the ordering of the operators involved.  For concreteness we will specify the operator ordering, although the choice ordering does not affect the results presented here.

First, we consider the canonical driven Jaynes-Cummings model
\begin{equation}\label{eq:jc}
\frac{\hat{H}}{\hbar} = -\Delta \hat{a}^{+}\hat{a}+\omega_{a}\hat{\sigma}^{+}_{-}\hat{\sigma}_{-}+g\left(\hat{a}+\hat{a}^{+}\right)\left(\hat{\sigma}_{-}+\hat{\sigma}^{+}_{-}\right) +E\left(\hat{a}+\hat{a}^{+}\right),
\end{equation}
where $\Delta=\omega_{d}-\omega_{c}$ is the detuning between the driving frequency $\omega_{d}$ and cavity frequency $\omega_{c}$.  The frequency for the two-level system is given by $\omega_{a}$, while $g$ represents the coupling strength between the atomic and cavity modes, and $E$ gives the amplitude of the external drive in units of frequency.  Note that we have not applied the rotating wave approximation to Eq.~(\ref{eq:jc}) as we want a numerical solution to the full Hamiltonian.  In addition, we consider a Liouvillian (\ref{eq:lindblad}) formed from cavity collapse operators $\sqrt{\kappa(1+n_{\rm th})}\hat{a}$ and $\sqrt{\kappa n_{\rm th}}\hat{a}^{+}$ corresponding to a thermal bath with an average thermal occupation number $n_{\rm th}$ at $\omega_{c}$, and qubit dissipation described by $\sqrt{\gamma(1+n_{\rm th})}\hat{\sigma}_{-}$, where $\hat{\sigma}_{-}$ is the qubit lowering operator.  The constants $\kappa$ and $\gamma$ characterize the energy decay rates for the cavity and qubit, respectively.  Here we are interested in the solution as the number of cavity states increases, and therefore fix the numerical parameters to be (in units of the cavity frequency $\omega_{c}$): $\Delta=0$, $\omega_{a}=1$, $g=0.25$, $E=1$, $\kappa=5\times 10^{-3}$, $\gamma=0.05$, and $n_{\rm th}=1$.  The modified Liouvillian $\tilde{\bm{\mathcal{L}}}$ representing Eq.~(\ref{eq:jc}) together with the collapse terms, is an example of a system that has zero elements along the diagonal.  As such, before using COLAMD ordering, we will always apply the weighted variant of MBM to first permute the diagonal to be zero-free. Here the operator ordering is given by $\hat{\sigma}_{-}=\hat{\sigma}_{-}\otimes \hat{\mathbb{I}}$ and $\hat{a}=\hat{\mathbb{I}}\otimes \hat{a}$

In addition, we consider a spin-chain consisting of $N$-spins where the first spin is being driven externally
\begin{eqnarray}\label{eq:spinchain}
\frac{\hat{H}}{\hbar}&=& -\frac{1}{2}\delta \hat{\sigma}^{(1)}_{z}-\frac{\Omega}{2} \hat{\sigma}^{(1)}_{x}-\frac{1}{2}\sum_{i=2}^{N}\omega_{n}\hat{\sigma}^{(n)}_{z} \\
&-& \frac{1}{2}\sum_{n=1}^{N-1}\left[J^{(n)}_{x}\hat{\sigma}^{(n)}_{x}\hat{\sigma}^{(n+1)}_{x}\right. \nonumber \\
&+&\left. J^{(n)}_{y}\hat{\sigma}^{(n)}_{y}\hat{\sigma}^{(n+1)}_{y}+J^{(n)}_{z}\hat{\sigma}^{(n)}_{z}\hat{\sigma}^{(n+1)}_{z}\right], \nonumber
\end{eqnarray}
where the frequency of each spin is denoted by $\omega_{n}$, $\delta=\omega_{d}-\omega_{1}$ is the detuning between the drive and resonant frequency for the first spin, $\Omega$ represents the driving strength, and the $J^{(n)}_{x,y,z}$ are the nearest neighbor coupling strengths.  For simplicity, we assume each spin has an identical frequency $\omega_{n}=2\pi$, and coupling strengths $J^{(n)}_{x,y,z}=0.1\times 2\pi$.  The driving amplitude is taken to be $\Omega=\omega_{1}/2$.  Finally, we also assume that each spin has a dephasing term given by $\sqrt{\gamma}\hat{\sigma}_{z}$, where $\gamma=0.01$.

\begin{figure*}[t]
\begin{center}
\includegraphics[width=13.0cm]{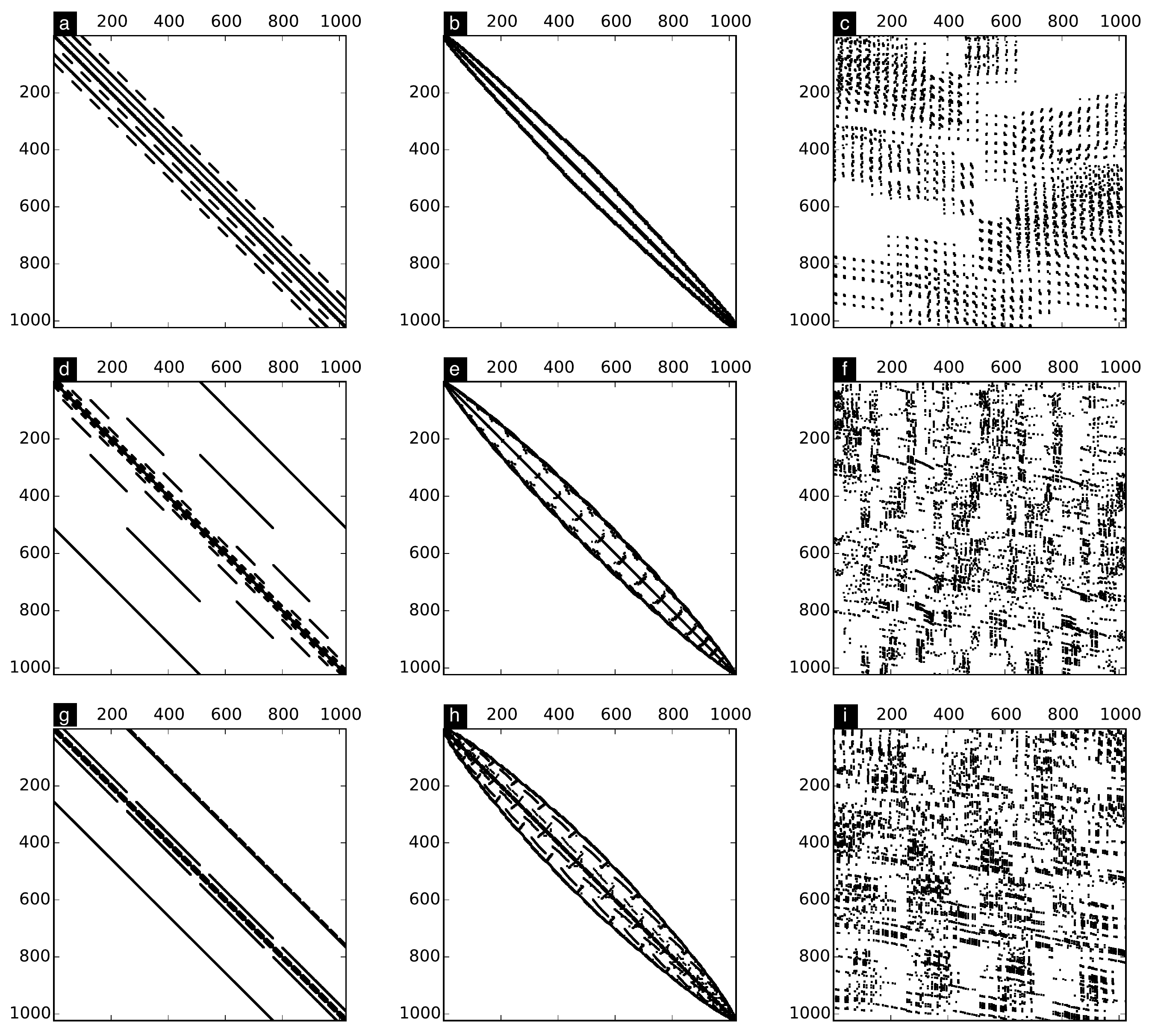}
\caption{a-c) Sparse matrix structure of the shifted Liouvillian super operator $\protect\overrightarrow{\bm{\mathcal{L}}}$ for the Jaynes-Cummings Hamiltonian (\ref{eq:jc}) with $N=16$ cavity states using natural, RCM, and COLAMD orderings.  The RCM bandwidth and profile is $137$ and $93329$, respectively, corresponding to a bandwidth reduction of $1.4$ and profile reduction of $1.7$. d-f) Sparse matrix representation of the shifted Liouvillian for the driven spin-chain (\ref{eq:spinchain}) consisting of $5$ spins in natural, RCM, and COLAMD ordering.  The RCM bandwidth and profile are $167$ and $117848$, respectively.  This corresponds to a bandwidth reduction of $6.1$, and a profile reduction of $4.9$. (g-i) Shifted Liouvillian for an optomechanical system (\ref{eq:opto}) comprised of $N_{c}=4$ cavity states and $N_{m}=8$ states in natural, RCM, and COLAMD ordering.  For RCM ordering, $B=229$ and $P=163176$ corresponding to bandwidth and profile reductions of $2.3$ and $2.5$, respectively.}
\label{fig:fig1}
\end{center}
\end{figure*}

Lastly, we look at an optomechanical Hamiltonian in a frame rotating at the driving frequency $\omega_{d}$ \cite{aspelmeyer:2014} 
\begin{equation}\label{eq:opto}
\frac{\hat{H}}{\hbar}=-\Delta\hat{a}^{+}\hat{a}+\omega_{m}\hat{b}^{+}\hat{b}+g_{0}(\hat{b}+\hat{b}^{+})\hat{a}^{+}\hat{a}+E\left(\hat{a}+\hat{a}^{+}\right),
\end{equation}
where $\Delta=\omega_{d}-\omega_{c}$, $\hat{a}$ and $\hat{b}$ are the annihilation operators for the cavity and mechanical modes, respectively, $\omega_{m}$ is the mechanical resonance frequency, and $g_{0}$ is the single-photon radiation pressure coupling strength.  In addition, we consider the experimentally relevant situation where the cavity mode is coupled to a thermal bath near zero temperature characterized by the collapse operator $\sqrt{\kappa}\hat{a}$, while the mechanical mode is in a nonzero thermal environment with associated collapse terms $\sqrt{\gamma(1+n_{\rm th})}\hat{b}$ and $\sqrt{\gamma n_{\rm th}}\hat{b}^{+}$.  The numerical parameters, in units of $\omega_{m}$ are: $\Delta=0$, $g_{0}=0.4$, $E=0.1$, $\kappa=0.3$, $\gamma=10^{-4}$, and $n_{\rm th}=1$,  while the operator ordering is $\hat{a}=\hat{a}\otimes \hat{\mathbb{I}}$ and $\hat{b}=\hat{\mathbb{I}}\otimes \hat{b}$.  Here we fix the number of cavity states to $N_{c}=4$ while varying the number of states $N_{m}$ for the mechanical resonator.

Given the reliance of each solution method on LU or iLU factorization, and the importance of the sparse matrix structure on the fill-in and stability of these solvers, in Fig.~\ref{fig:fig1} we plot the matrix structures (non-zero elements) for the shifted Liouvillian $\overrightarrow{\bm{\mathcal{L}}}$ used in the inverse-power method for the three representative systems using natural, RCM, and COLAMD ordering.  In addition, we give the bandwidth and profile reduction factors, $B(\overrightarrow{\bm{\mathcal{L}}})/B(\overrightarrow{\bm{\mathcal{L}}}_{\rm RCM})$ and $P(\overrightarrow{\bm{\mathcal{L}}})/P(\overrightarrow{\bm{\mathcal{L}}}_{\rm RCM})$ respectively, when using RCM ordering. In all three cases, the matrix structure is sufficiently close to symmetric such that RCM ordering reduces both the bandwidth and profile for each of the shifted system Liouvillian operators.  As such, we would expect that LU factorization using this ordering results in fewer nonzero elements, and thus outperforms the naturally ordered Liouvillian in terms of both runtime and memory consumption.  Similar reasoning applies in the case of iLU factorization, however the stability of the approximate inverse is also critical, and both the structure and numerical values of the elements play a crucial role. The comparison to COLAMD reordering is not as straightforward, and must be evaluated by performing the numerical factorization.

\begin{figure*}[t]
\begin{center}
\includegraphics[width=16.0cm]{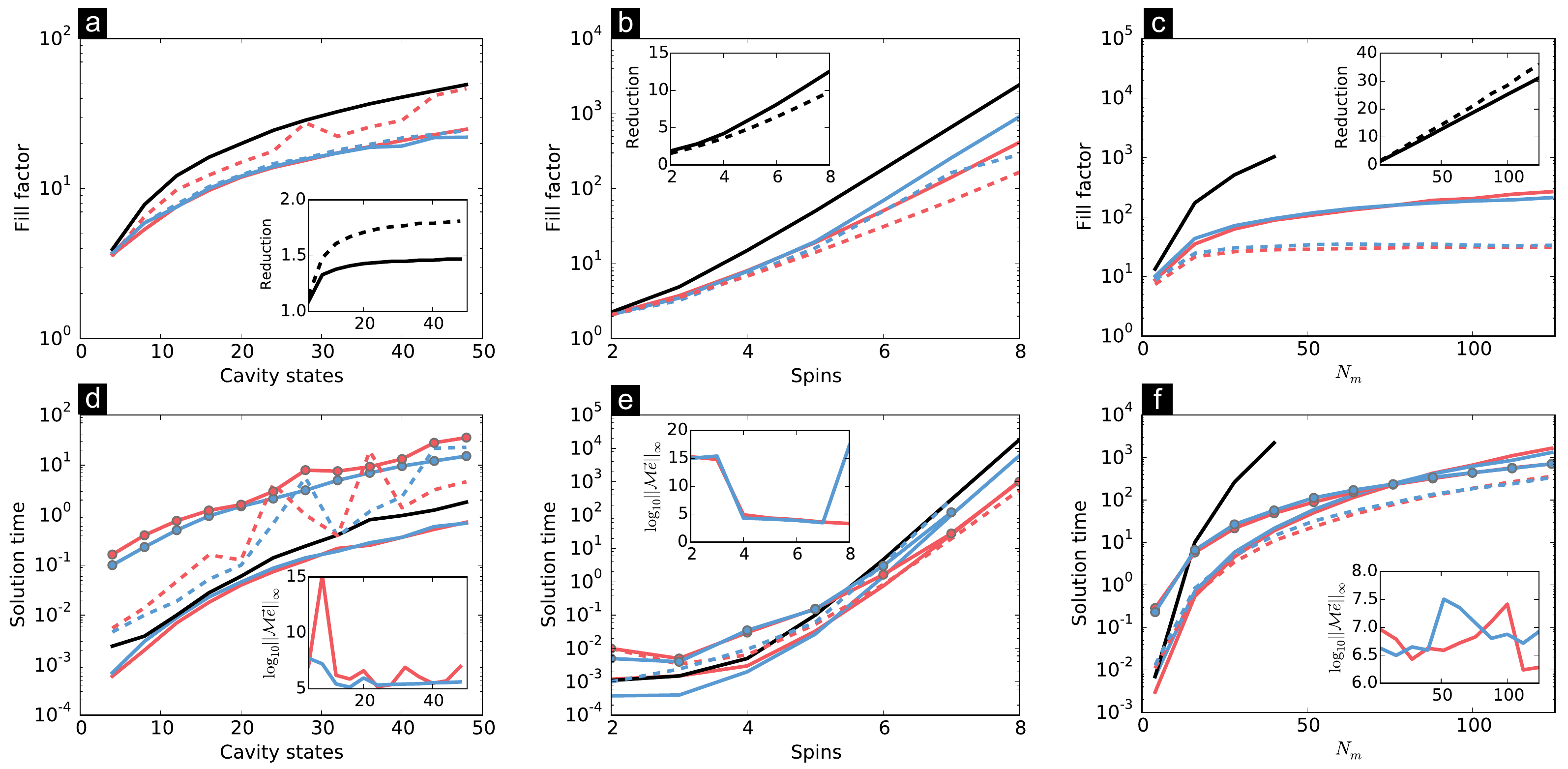}
\caption{(a) Fill factors for the complete LU decomposition of the shifted Liouvillian $\protect\overrightarrow{\bm{\mathcal{L}}}$ used in the inverse power method for a Jaynes-Cummings system as a function of the number of cavity states using RCM (solid-red), COLAMD (solid-blue), and natural (black) orderings.  iLU factorization fill factors for RCM (dashed-red) and COLAMD (dashed-blue) orderings are also presented.   Inset figure shows the bandwidth (solid) and profile (dashed) reduction factors when using RCM ordering. (b) Fill-factors for the driven spin-chain for the same methods used in (a). (c) Fill-factors for the shifted optomechanical Liouvillian as a function of the number of mechanical Fock states $N_{m}$. The natural ordering (black) cannot be computed beyond $N_{m}=40$ due to memory limitations. (d) Solution times for finding the steady state density matrix to a tolerance of $10^{-14}$, using the inverse-power method with RCM (solid-red), COLAMD (solid-blue), or natural (black) ordering.  In addition, solutions using iterative GMRES (solid-dots) or BICGSTAB (dashed) solvers for LU factorization based on RCM (red) and COLAMD (blue) orderings are included.  The inset shows the the approximate condition number of the iLU factorization $\mathcal{M}$ based on these orderings. (e) Solution times for the driven spin-chain model.  Here, both the BICGSTAB and GMRES iterative methods based on COLAMD ordering fail for $N=8$ spins due to a failure to converge to the desired tolerance in $1000$ iterations.  (f) Timing results for the shifted optomechanical Liouvillian.  Memory restrictions limit the solution using natural matrix ordering to $N_{m}\le 40$.}
\label{fig:fig2}
\end{center}
\end{figure*}

In Fig.~\ref{fig:fig2} we present the fill factors and solution times for each system using the inverse-power method, where the factorization of $\overrightarrow{\bm{\mathcal{L}}}$ is performed using LU factorization using natural, RCM, or COLAMD ordering, and the doubly-iterative inverse-power solver based on  iLU factorization employing RCM and COLAMD permutations using both GMRES and BiCGSTAB solvers.  In addition, the bandwidth and profile reduction factors, and in the case of iLU factorization condition estimates for the approximate inverse $\bm{\mathcal{M}}$, are also given. In all cases, we set the solution tolerance for the inverse-power solver to $10^{-14}$.  All simulations were carried out using the QuTiP framework \cite{qutipweb,qutip:2012,qutip:2013}, where we have used the default setting for the various solvers save for the allowed fill-in $p$ that we set to $p=300$ so as to accommodate all of the simulation results using a single set of parameters.  

Figure~\ref{fig:fig2} highlights the benefit of both RCM and COLAMD ordering of the shifted Liouvillian operator in LU factorization for each system.  For the largest Hilbert space dimensions considered here, the reduction in fill-in corresponds to a factor of five or more savings in terms of the memory requirements for direct LU factorization.  iLU factorizations reduce these memory requirements even further, by nearly another order of magnitude, except in case of the Jaynes-Cummings model, where  RCM ordering gives a larger fill factor even though iLU factorization should result in fewer nonzero elements.  COLAMD iLU factorization also does no better than the full LU factorizations.  This suggests that the default values used in the creation of the preconditioner are ill-suited for this particular system.   In contrast, RCM ordering of the spin-chain Liouvillian reduces the memory footprint of factorization by over a factor of twenty compared to natural ordering, and a factor of two reduction when compared with COLAMD iLU factorization using the same parameters.

\begin{figure*}[t]
\begin{center}
\includegraphics[width=13.0cm]{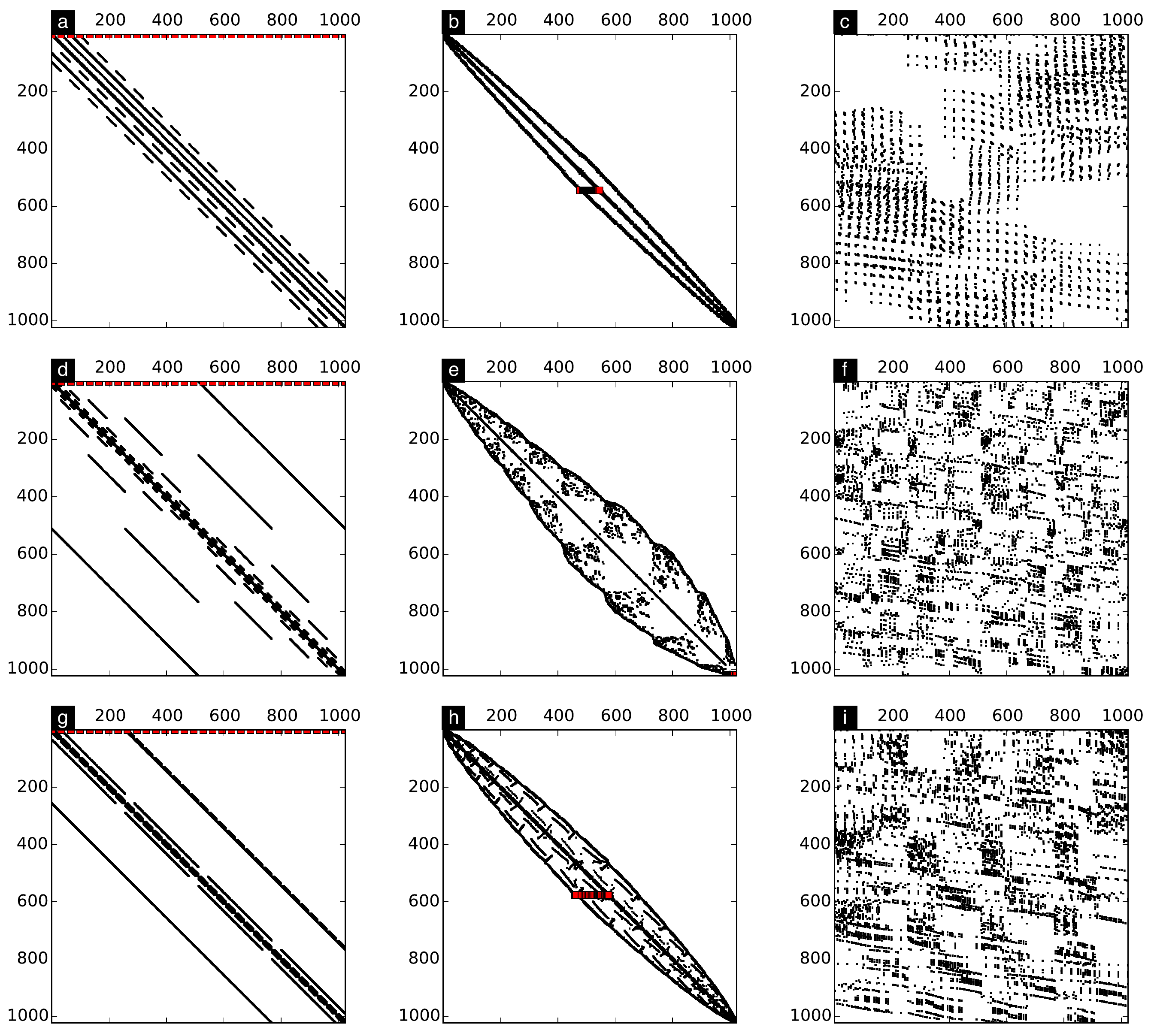}
\caption{(a-c) Sparse matrix structure of the modified Liouvillian $\tilde{\bm{\mathcal{L}}}$ for the Jaynes-Cummings model under natural, RCM, and COLAMD ordering.  The location of the elements for $\bm{\mathcal{T}}$ (red) are given for both the natural and RCM orderings and are enlarged for clarity. Under RCM ordering, the modified Liouvillian has bandwidth $B=139$ and profile $P=93300$, giving bandwidth and profile reduction factors of $8.0$ and $1.7$, respectively. (d-f) Matrix structure for the driven spin-chain in natural, RCM, and COLAMD order.  Under RCM ordering, the elements of $\bm{\mathcal{T}}$ have been permuted from first to last row giving $B=405$ and $P=248489$, corresponding to a bandwidth reduction of $3.8$ and profile reduction factor $2.35$.  (g-i) The modified optomechanical Liouvillian after natural, RCM ($B=239$, $P=163299$), and COLAMD permutations.  The RCM bandwidth and profile reduction factors are $5.4$ and $2.5$, respectively}
\label{fig:fig3}
\end{center}
\end{figure*}
Given that the solution time is proportional to the number of nonzero elements in the LU or iLU factors, the solution times for each method are largely in step with the fill factors generated for each reordering.  The biggest gains in performance occur for the driven spin chain where LU factorization using RCM reordering outperforms COLAMD by nearly a factor of six, and natural ordering by over an order of magnitude.  For iterative solutions, the solution times when using the BiCGSTAB method are generally lower than those using the GMRES solver for each ordering indicating that the BiCGSTAB method achieves convergence in a relatively few number of iterations.  For the driven spin-chain with $N=8$ spins, both GMRES and BiCGSTAB iterative methods fail to converge to the requested tolerance value when using COLAMD ordering.  This lack of converge can be understood by examining the condition number estimate for the approximate inverse, Fig.~\ref{fig:fig2}e, that suggests that the computed inverse is too ill-conditioned to reach the requested level of accuracy.  The relatively lower condition number achieved using RCM ordering is in line with earlier investigations \cite{nation:2015} that suggest that the stability provided by this ordering plays a key role in the success of iterative methods for open quantum systems.  Note that at smaller system sizes, the conditioning of the approximate inverse is not as crucial as in the large dimensional case as the system Liouvillian itself is better conditioned for relatively smaller Hilbert space sizes.

Turning now to direct solutions to Eq.~(\ref{eq:modified}), in Fig.~\ref{fig:fig3} we plot the matrix structure for the modified Liouvillian $\tilde{\bm{\mathcal{L}}}$ for each example system, and in particular, highlight the location of the matrix elements corresponding to the trace preserving matrix $\bm{\mathcal{T}}$ that gives the large upper bandwidth in the naturally ordered Liouvillian operators.  In addition we give the bandwidth $B(\bm{\tilde\mathcal{L}})/B(\bm{\tilde\mathcal{L}}_{\rm RCM})$ and profile $P(\bm{\tilde\mathcal{L}})/P(\bm{\tilde\mathcal{L}}_{\rm RCM})$ reduction factors when using RCM ordering for the modified Liouvillian.  For both the Jaynes-Cummings and optomechanical modified Liouvillians, the total bandwidth and profile after RCM permutation are nearly the same as those for the shifted Liouvillian, and the introduction of $\bm{\mathcal{T}}$ should not greatly affect the factorization properties of these systems.  In stark contrast, the bandwidth of the driven spin-chain modified Liouvillian is nearly four times greater than the shifted Liouvillian, Fig.~\ref{fig:fig1}e, as the need to accommodate the addition of the trace preserving matrix, and the associated loss of symmetry,  reduces the effectiveness of RCM ordering.  In this case, we expect the solution time to be markedly longer than that of the inverse-power method.
\begin{figure*}[t]
\begin{center}
\includegraphics[width=16.0cm]{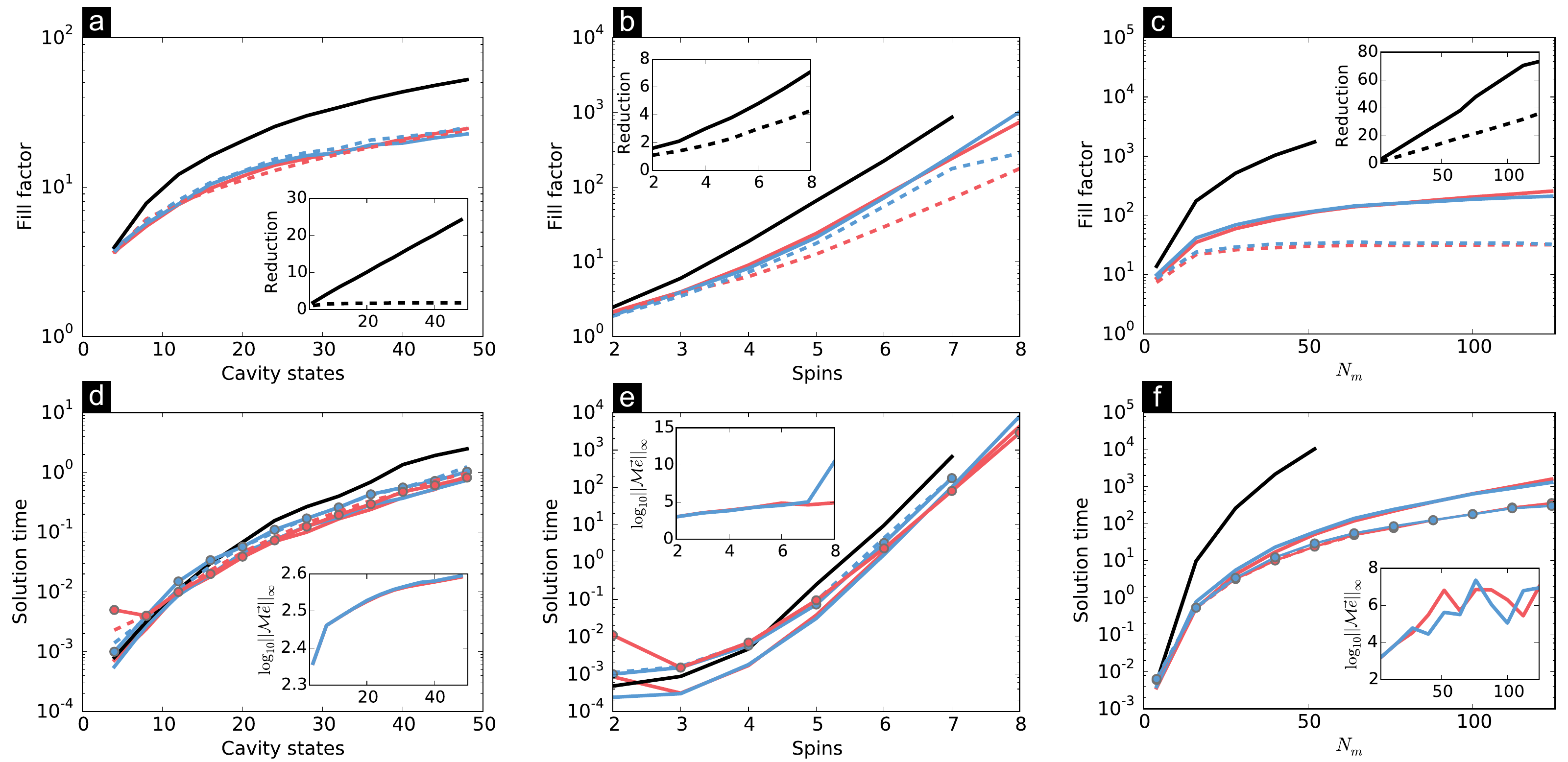}
\caption{(a) Fill factors for the complete LU decomposition of the modified Liouvillian $\tilde{\bm{\mathcal{L}}}$ for the Jaynes-Cummings system as a function of the number of cavity states using RCM (solid-red), COLAMD (solid-blue), and natural (black) orderings.  iLU factorization fill for RCM (dashed-red) and COLAMD (dashed-blue) orderings are also given.   The inset gives the bandwidth (solid) and profile (dashed) reduction factors when using RCM ordering. (b) Fill factors for the modified Liouvillian corresponding to the driven spin chain as a function of the number of spins for the various factorization methods. (c) Fill factors for an optomechanical system with $N_{c}=4$ as the number of mechanical states is varied.  Here, the natural ordering fill factor becomes larger than the available memeory after $N_{m}>50$. (d) Solution times for finding the steady-state density matrix to a tolerance of $10^{-14}$, using direct LU factorization with RCM (solid-red), COLAMD (solid-blue), or natural (black) ordering of  $\tilde{\bm{\mathcal{L}}}$. Solutions using iLU factorization based on RCM (red) and COLAMD (blue) for iterative GMRES (solid-dots) or BICGSTAB (dashed) solvers are also presented.  The inset shows the the approximate condition number of the iLU factorization $\mathcal{M}$ based on these orderings. (e) Solution times for the driven spin-chain model.  Here, both the BICGSTAB and GMRES iterative methods based on COLAMD ordering fail for $N=8$ spins due to a failure to converge to the desired tolerance in $1000$ iterations. (f) Timing results for the modified optomechanical Liouvillian.  Memory restrictions limit the solution using natural matrix ordering to $N_{m}\le 50$.  For all simulations, the weighting factor $w$ in Eq.~(\ref{eq:linear}) is set to the average of the diagonal elements in the modified Liouvillian.}
\label{fig:fig4}
\end{center}
\end{figure*}

Finally in Fig.~\ref{fig:fig4} we present the fill-in and solution times for solving the steady-state density matrix calculated from the direct LU and iterative iLU factorizations of the modified Liouvillian (\ref{eq:modified}) for the three representative systems.  The results for fill-in closely parallel those in Fig.~\ref{fig:fig2}, however the fill-in for the iLU factorization using RCM ordering of the Jaynes-Cummings Liouvillian is more stable when varying the number of cavity states.  The benefit of direct LU factorization using RCM reordering for the driven spin-chain is also reduced due to the larger bandwidth, however it is still a factor of two lower than that of COLAMD.  The RCM LU and iLU factorization for the optomechanical system still perform well due to the negligible increase in bandwidth and fill-in compared to those used in the inverse-power method.

Once again, the solution times when using complete LU factorization are directly determined by the fill-in, and are longer than those of the inverse-power method as the addition of the matrix $\bm{\mathcal{T}}$ increases the total number of nonzero elements in the modified Liouvillian. that, in combination with higher fill-ins, gives an increased runtime.  The large increase in bandwidth seen for the driven spin-chain in turn leads to longer solution times that, for $N=8$, are four times longer than the inverse-power methods.  In cases such as this, it is advantageous to compare the RCM bandwidth for the shifted and modified Liouvillians, and pick the appropriate solution method based on the result.  Fortunately, calculating the RCM ordering and bandwidth takes only a fraction of the total computation time, and can readily be checked before performing the more costly factorization step.

For iterative iLU methods, the results presented in Fig.~\ref{fig:fig4} indicate that the preferred Krylov solver, in terms of runtime, is now the GMRES method.  This is an indication that the number of iterations needed to achieve our tolerance goal using the BiCGSTAB method, is computationally more intensive than building the $m=20$ (the default value in QuTiP) Krylov subspace for the GMRES solver and using fewer iterations.   In addition, the BiCGSTAB method completely fails for the optomechanical system as the residual vector computed during the iteration process becomes orthogonal to the original residual, leading to a break down in the algorithm \cite{barrett:1993}.  The iterative techniques using iLU factorization based on COLAMD ordering once again fail for the driven spin-chain when $N=8$ due to the poor conditioning of the approximate inverse.  Although not seen here, the condition number of $\bm{\mathcal{M}}$ also becomes important for large dimensional optomechanical systems where iLU factorization was shown to fail for COLAMD ordering, but converged to machine precision tolerance when using RCM \cite{nation:2015}.  In contrast, RCM ordering is found to lead to convergence to the requested tolerance value for the GMRES method in all cases, and is now the fastest solution method for the spin-chain and optomechanical systems at large Hilbert dimensions.  Moreover, the memory savings obtained from iLU factorization using RCM is close to an order of magnitude smaller than full LU factorization using any reordering.  In contrast, the direct solution of the Jaynes-Cummings model using COLAMD is still the optimal method.  Here, the dimensionally of the underlying Hilbert space is never large enough for iterative methods and matrix permutation to overcome the intrinsically fast performance of direct LU factorization on small systems that is also seen in the other two systems.

\section{Conclusion}\label{sec:conclusion}
We have examined the use of available numerical solution methods and matrix reordering strategies for solving for the steady-state density matrix for several common time-independent systems found in quantum optics and related sub-disciplines.  In addition, we have introduced a doubly-iterative inverse-power method, making use of the fast, usually single step, convergence of the inverse-power method to replace full LU factorization with an iLU decomposition.   These solution techniques were tested on several standard quantum optical systems were it is found that iterative methods based on RCM reordering of the system Liouvillian outperform techniques based on direct LU factorization for large Hilbert space dimensions.  Moreover, iterative solvers using COLAMD ordering fail at large dimensions due to poor conditioning of the approximate inverse, while those based on RCM remain stable.  Provided that the RCM bandwidth of the modified Liouvillian $\tilde{\bm{\mathcal{L}}}$ is nearly the same as that of the shifted Liouvillian $\overrightarrow{\bm{\mathcal{L}}}$ then preconditioned GMRES solvers using $\tilde{\bm{\mathcal{L}}}$ should be the first choice of iterative method as one does not need to worry about eigenvalue conditioning.  If instead, bandwidth reduction becomes hampered, due to the introduction of the trace preserving matrix $\bm{\mathcal{T}}$, then the doubly-iterative inverse-power method can be employed.  Regardless of which iterative method is used, finding an appropriate drop-tolerance and fill-factor for the approximate inverse must still be done on a trial and error basis, although the default values used in QuTiP are a good starting point. 

In contrast to iterative methods, a clear understanding of the optimal matrix permutation method for direct LU factorization is still lacking.  Indeed, there is no a priori information by which to judge whether RCM or COLAMD will be the best choice in terms of fill-in, or perhaps if some other as yet unstudied reordering will be most efficient.  Although these orderings give fill-ins that are usually on par with each other, for large quantum systems, the choice of permutation can mean the difference between finding a solution and running into memory limitations.  Therefore, understanding how the Liouvillian matrix structure affects the fill-in in each of these methods is a critical area of future research.  Likewise, exploring additional iLU reorderings based on BFS (like RCM), or the related depth-first search, may give enhanced bandwidth and profile reduction while maintaining good conditioning of the approximate inverse \cite{benzi:1999,bridson:2000}.  As with the results presented here, this line of endeavors will help to extend the applicability of classical solution methods to the steady-state density matrix to ever larger quantum systems in absence of a quantum computer.

All of the numerical algorithms presented in this work are freely available in the QuTiP: Quantum Toolbox in Python library \cite{qutipweb,qutip:2012,qutip:2013}.

\section*{Acknowledgments}
P.D.N was supported by startup funding from Korea University.

\section*{References}
\bibliographystyle{iopart-num}
\bibliography{text}
\end{document}